\documentclass[3p,times]{elsarticle}

\usepackage{ecrc}

\usepackage{epstopdf}

\volume{00}

\firstpage{1}

\journalname{Nuclear Physics A}

\runauth{Author1 et al.}

\jid{nupha}

\jnltitlelogo{Nuclear Physics A}

\usepackage{graphicx}
\usepackage{epsfig,bm,dcolumn}
\usepackage{amsmath,amssymb}

\begin{document}

\begin{frontmatter}



\title{$\Upsilon$ Production in Heavy Ion Collisions at LHC}

\author{Kai Zhou$^1$, Nu Xu$^{2,3}$, Pengfei Zhuang$^1$}
\address{$^1$Physics Department, Tsinghua University and Collaborative Innovation Center of Quantum Matter, Beijing 100084, China\\
             $^2$Nuclear Science Division, Lawrence Berkeley National Laboratory, Berkeley, CA 94720, USA\\
             $^3$Key Laboratory of Quark and Lepton Physics (MOE) and Institute of Particle Physics, Central China Normal University, Wuhan 430079, China}




\begin{abstract}
We investigate the $\Upsilon$ production in heavy ion collisions at LHC energy in the frame of a dynamical transport
approach. Both the initial production and in-medium regeneration and both the cold and hot nuclear matter effects are included in the calculations. In comparison with the ground state $\Upsilon(1s)$, the excited state $\Upsilon(2s)$ is much more sensitive to the heavy quark potential at finite temperature.
\end{abstract}


\end{frontmatter}


\indent

\indent

In comparison with the study of charmonium suppression~\cite{matsui,spsjpsi,rhicjpsi,alicejpsi}, the bottomonium production is expected to offer a cleaner probe to the hot medium created in the early stage of relativistic heavy ion collisions. Since $\Upsilon$s are more tightly bound states of heavy quarks, the theoretical calculations on the production and dissociation cross sections via perturbation expansion~\cite{bhanot} and dissociation temperatures through Schr\"odinger equation~\cite{satz} become more solid and can provide more reliable predictions. Different from the calculations at RHIC energy where the rare bottom quarks in the fireball leads to negligible $\Upsilon$ regeneration and the initial production becomes the only source~\cite{grandchamp,liu1,strickland}, the regeneration plays an important role at LHC energy, due to the increased bottom quark number and the strongly suppressed initial contribution.

In this paper we investigate $\Upsilon$ production in Pb+Pb collisions at LHC in the frame of our transport approach~\cite{yan} which describes well the yield and especially the transverse momentum distribution of $J/\psi$ at LHC~\cite{zhou}. Since the dynamical effect of the hot medium on the quarkonium dissociation and regeneration cross sections is through the heavy quark potential at finite temperature, we will focus on the potential dependence of different $\Upsilon$ states.

Introducing the prompt time $\tau=\sqrt{t^{2}-z^{2}}$, space-time rapidity
$\eta=\frac{1}{2}\ln\frac{t+z}{t-z}$ , momentum rapidity $y=\frac{1}{2}\ln\frac{E+p_{z}}{E-p_{z}}$ and transverse energy $E_{T}=\sqrt{E^{2}-p_{z}^{2}}$ to replace $t,z,p_{z}$ and
$E=\sqrt{m^{2}+{\bf p}^{2}}$, the $\Upsilon$ distribution function $f_\Upsilon(\eta,{\bf x}_T,y,{\bf p}_T,\tau|{\bf b})$ in heavy ion collisions with impact parameter ${\bf b}$ is controlled by the Boltzmann equation~\cite{liu2}
\begin{equation}
\left[\cosh(y-\eta){\frac{\partial}{\partial\tau}}+{\frac{\sinh(y-\eta)}{\tau}}{\frac{\partial}{\partial
\eta}}+{\bf v}_T\cdot\nabla_T\right]f_\Upsilon
=- \alpha_\Upsilon f_\Upsilon+\beta_\Upsilon,
\label{trans2}
\end{equation}
where ${\bf v}_{T}={\bf p}_{T}/E_{T}$ is the $\Upsilon$ transverse velocity.

Considering the gluon dissociation $\Upsilon+g\to b+\bar{b}$ as the dominant dissociation
process in the hot and dense quark matter, the loss term $\alpha_{\Upsilon}$ can be calculated by the gluon momentum
integration of the dissociation cross section $\sigma_{g\Upsilon}$ multiplied by the thermal gluon distribution $f_{g}$ and the flux factor $F_{g\Upsilon}$,
\begin{equation}
\label{loss}
\alpha_\Upsilon=\frac{1}{2E_T}\int
{d^3{\bf k}\over (2\pi)^3
2E_g}\sigma_{g\Upsilon}({\bf p},{\bf k},T)4F_{g\Upsilon}({\bf p},{\bf k})f_g({\bf k},T),
\end{equation}
where $E_{g}$ is the gluon energy. The dissociation cross section in vacuum can be derived through the
operator product expansion with perturbative Coulomb wave function\cite{bhanot}, and its temperature dependence can be estimated by taking the geometrical relation between the cross section and the average size of the bottomonium state,
$\sigma_{g\Upsilon}({\bf p},{\bf k},T)=\sigma_{g\Upsilon}({\bf p},{\bf k},0)\langle r_{\Upsilon}^{2}\rangle(T)/\langle r_{\Upsilon}^{2}\rangle(0)$ with the averaged radius
square $\langle r_{\Upsilon}^{2}\rangle$ from the potential model~\cite{satz}. The local temperature $T({\bf x},t|{\bf b})$ and the fluid velocity
$u_{\mu}({\bf x},t|{\bf b})$ of the medium which appear in the gluon distribution function $f_g$ are determined by the relativistic hydrodynamical equations~\cite{yan} which describes the space-time evolution of the medium.

The gain term $\beta_\Upsilon$ due to the in-medium regeneration is connected to the loss term $\alpha_\Upsilon$ via detailed banlance between the gluon dissociation process and
it's inverse process. Considering the experimentally observed large B quark quench factor~\cite{cms1}, we take as a first order approximation a kinetically thermalized momentum spectrum for the bottom distribution $f_b$. Neglecting the creation and annihilation of bottom-antibottom pairs inside
the medium, the spacial density of bottom (antibottom) number $\rho_b({\bf x},t|{\bf b})=\int d^3{\bf q}/(2\pi)^3f_b({\bf
x},{\bf q},t|{\bf b})$ satisfies the conservation law $\partial_\mu\left(\rho_b u^\mu\right)=0$ with the initial density determined by the nuclear geometry $\rho_b({\bf x},\tau_0|{\bf b})=T_A({\bf x}_T)T_B({\bf x}_T-{\bf b})\cosh\eta/\tau_0 d\sigma^{pp}_{b\bar b}/ d\eta$, where $T_A$ and $T_B$ are the thickness functions at transverse
coordinate ${\bf x}_T$, and
$d\sigma^{pp}_{b\bar b}/d\eta$ is the rapidity distribution of the bottom quark 
production cross section in p+p collisions.

Considering the feed-down contribution from the excited states to the ground state, we should take into account transport
equations for all the bottomonium states, when we calculate the final state $\Upsilon(1s)$ distribution. For simplicity, we treat $\Upsilon(3S)$ the same as $\Upsilon(2S)$ and
$\Upsilon(2P)$ the same as $\Upsilon(1P)$. Then from the experimental measurement~\cite{affolder}, the contribution from the directly produced $\Upsilon(1s)$s to the finally observed $\Upsilon(1s)$s is only $51\%$, and the decay contribution is $37\%$ from $\Upsilon(1p)$ and $12\%$ from $\Upsilon(2S)$~\cite{liu1}.

From PYTHIA simulation~\cite{pythia} for p+p collisions, we take the initial $\Upsilon$ distribution at time $\tau_0$ to be the following power law
\begin{equation}
\label{ppjpsi}
{d^2\sigma^{pp}_\Upsilon\over dy p_Tdp_T}=\frac{2(n-1)}{(n-2)\langle p_T^2\rangle_{pp}}\left(1+{\frac{p_T^2}{(n-2)\langle p_T^2\rangle_{pp}}}\right)^{-n}{\frac{d\sigma^{pp}_\Upsilon}{dy}}
\end{equation}
with $n=3$ and $\langle p_T^2\rangle_{pp}(y)=20(\text{GeV/c})^2(1-y^2/y_{max}^2)$, where $y_{max}=\text{arccosh}(\sqrt{s_{NN}}/(2m_{\Upsilon}))$ is the most forward rapidity of $\Upsilon$, and the expression for the rapidity density $(d\sigma^{pp}_\Upsilon/dy)/(d\sigma^{pp}_\Upsilon/dy|_{y=0})=e^{-y^2/0.33y_{max}^2}$ with $d\sigma^{pp}_\Upsilon/dy|_{y=0}=40$ nb gives $d\sigma_{\Upsilon}^{pp}/dy|_{y=3.25}=16$ nb which is consistent with the LHCb measurement~\cite{lhcb}. For the bottom quark production cross section, we take $d\sigma_{b\bar b}^{pp}/dy|_{y=0}=20\ \mu b$ estimated from FONLL\cite{fonll} and $\sigma_{b\bar b}^{pp}/dy|_{y=3.25}=7\ \mu b$ from FONLL scaling. To simply incorporate the shadowing effect for bottom quark, we take a $20\%$ reduction for the bottom quark number from the estimation with EKS98 shadowing evolution~\cite{eks98}.

\begin{figure}[htb]
\centering
\includegraphics[width=0.45\textwidth]{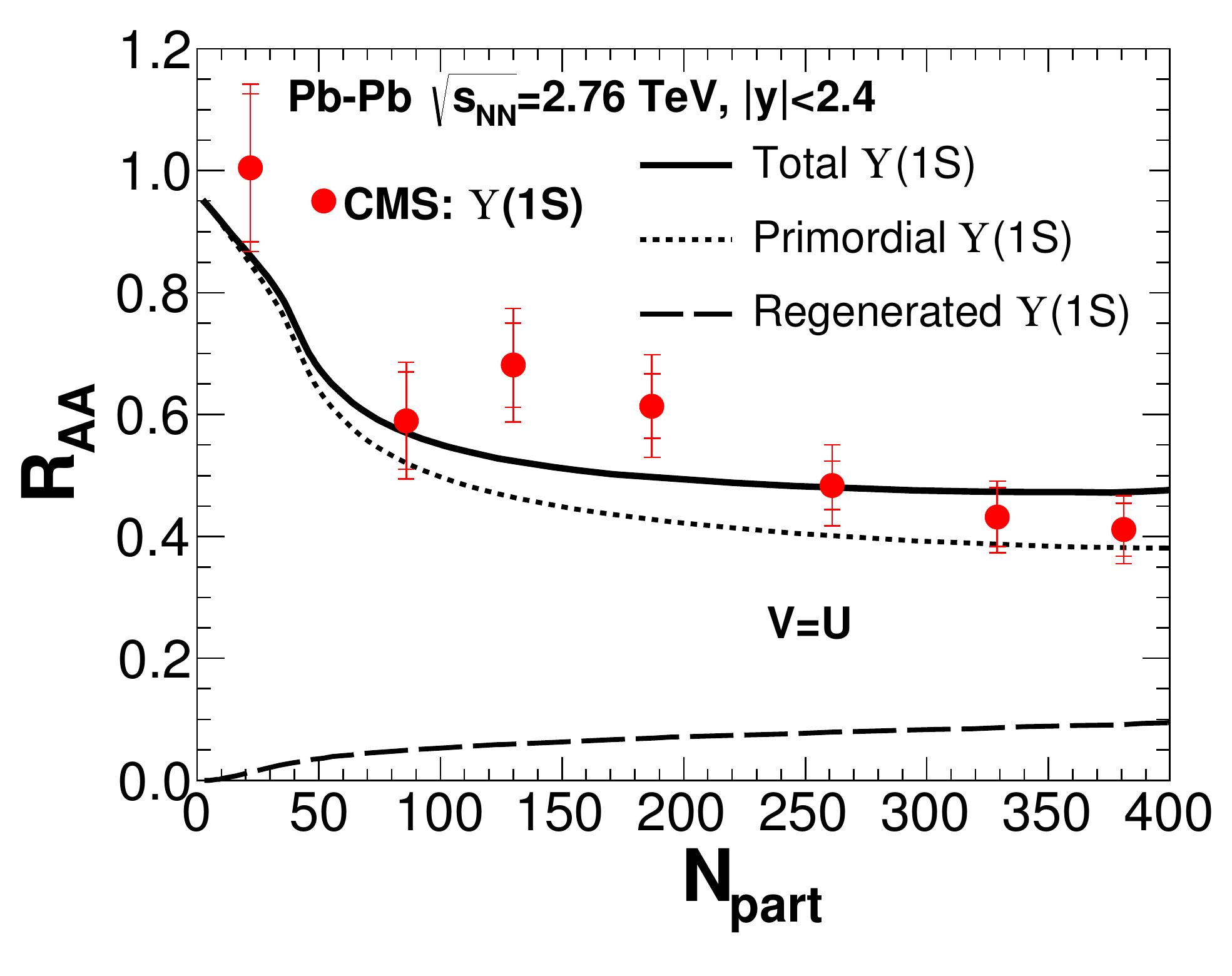}
\includegraphics[width=0.45\textwidth]{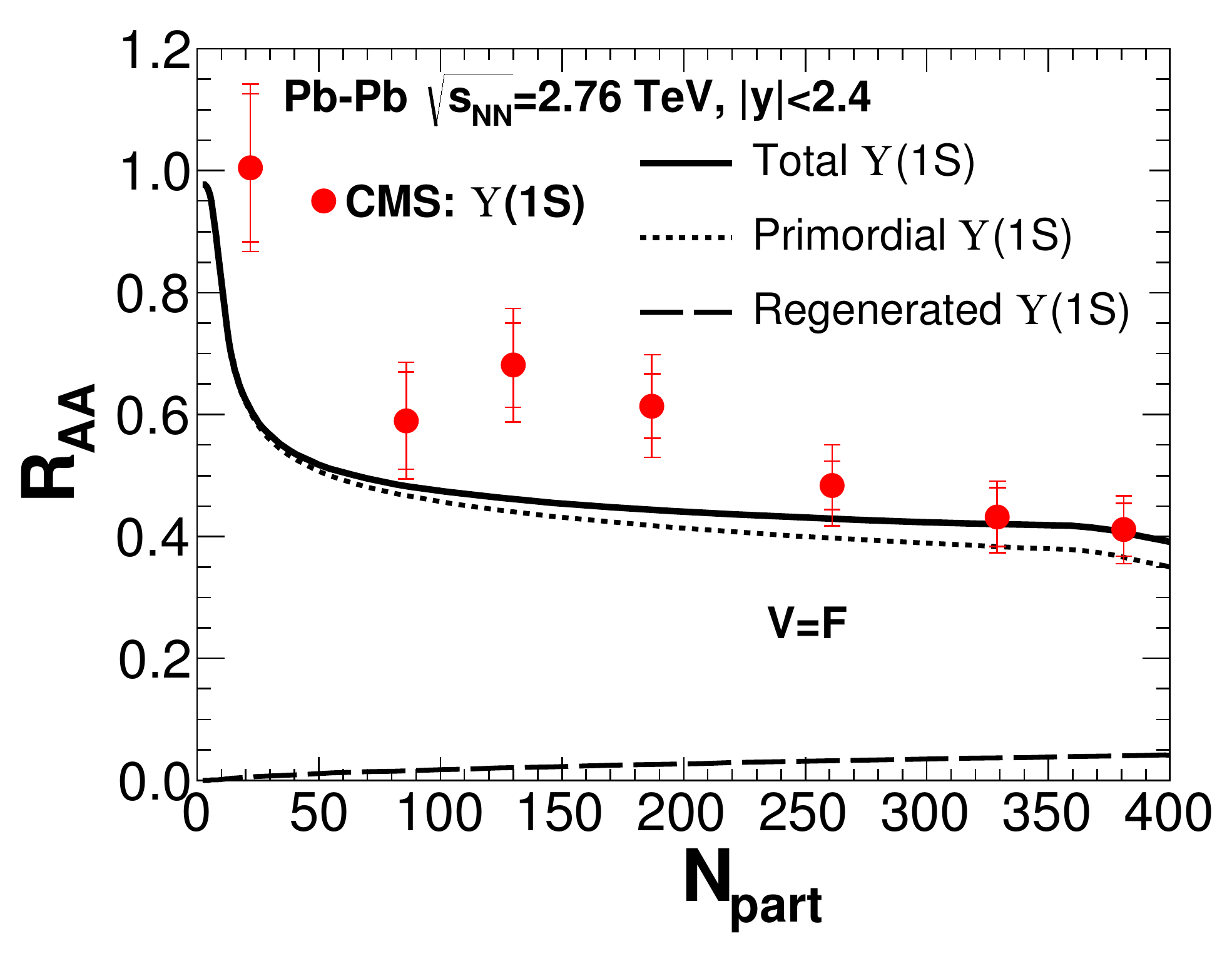}
\caption{The nuclear modification factor for $\Upsilon(1s)$ as a
function of centrality in 2.76 TeV Pb+Pb collisions at mid rapidity. The data are from CMS Collaboration~\cite{cms2}, and the dotted, dashed and solid lines are the transport calculations with only initial production, only regeneration and both in the two limits of $V=U$ (left panel) and $V=F$ (right panel). }
\label{fig1}
\end{figure}

With the obtained distribution function $f_\Upsilon$ in phase space at time $\tau\to\infty$, we can calculate the $\Upsilon$ yield and momentum distribution. As we have seen, the temperature dependence of the dissociation and regeneration cross sections comes from the average size $\langle r_\Upsilon^2\rangle (T)$ which is determined by the heavy quark potential $V$ at finite temperature $T$. From the lattice QCD~\cite{lattice}, we can extract the heavy quark free energy $F$ and the internal energy $U=F+TS=F-T\partial F/\partial T$. The potential should be in between $F$ and $U$. To see the maximum dependence of the $\Upsilon$ production on the potential, we only consider the two limits, $V=F$ and $V=U$. We start with the nuclear modification factor $R_{AA}=N_{AA}/\left(N_{coll} N_{pp}\right)$ as a function of the number of participants $N_{part}$, where $N_{pp}$ and $N_{AA}$ are respectively the numbers of measured $\Upsilon$s in p+p and A+A collisions, and $N_{coll}$ is the number of nucleon-nucleon
collisions at fixed $N_{part}$. The model calculation for 2.76 TeV Pb+Pb collisions and the comparison with the CMS data~\cite{cms2} for the ground state $\Upsilon(1s)$ are shown in
Fig.\ref{fig1}. In the limit of $V=U$ (left panel of Fig.\ref{fig1}) which leads to the maximum dissociation temperature $T_d(\Upsilon)/T_c=4.1, 1.8, 1.6$ for $\Upsilon(1S), \Upsilon(1P)$ and $\Upsilon(2S)$, the fireball temperature determined by the hydrodynamic equations is much lower than $T_d(\Upsilon(1S))$ but much higher than $T_d(\Upsilon(1p))$ and $T_d(\Upsilon(2s))$, and therefore there is only a weak dissociation for the ground state but very strong suppression for the excited states. If we consider only the initial production (dotted lines) and simply assume that all $\Upsilon(1s)$s can survive and all the excited states are dissociated, the nuclear modification factor for $\Upsilon(1S)$ would be 0.51 estimated from the decay fractions. When the regeneration which is still small at LHC energy is taken into account (dashed lines), the total result (solid lines) agrees well with data. In the other limit of $V=F$ (right panel of Fig.\ref{fig1}) which results in the minimum dissociation temperature $T_
d(\Upsilon)/T_c=3.0, 1.1, 1.0$ for $\Upsilon(1S), \Upsilon(1P)$ and $\Upsilon(2S)$, the above relation between the fireball temperature and dissociation temperature is still qualitatively valid for the $\Upsilon$ states, and the $\Upsilon(1s)$ yield is not sensitive to the potential.

\begin{figure}[htb]
\centering
\includegraphics[width=0.45\textwidth]{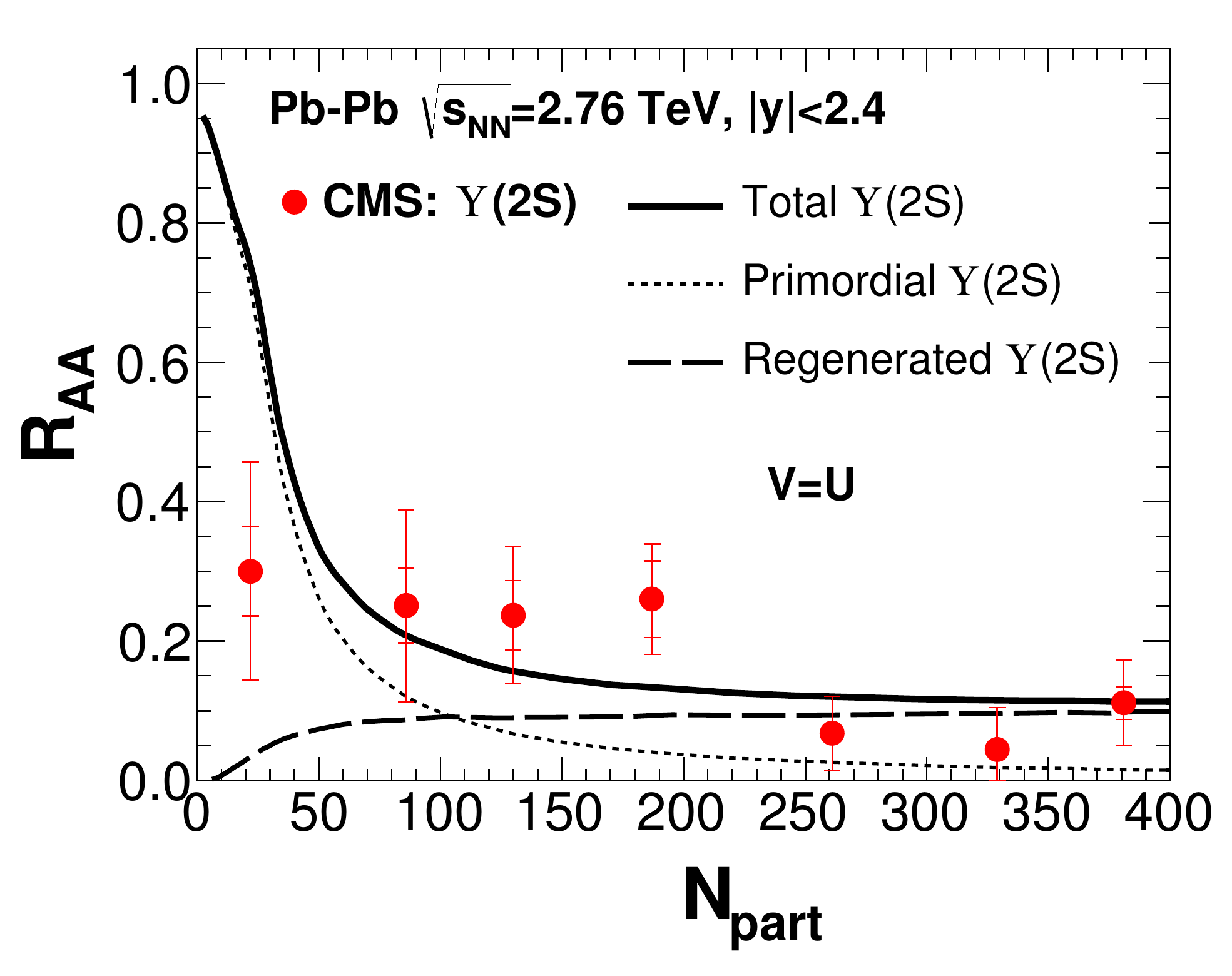}
\includegraphics[width=0.45\textwidth]{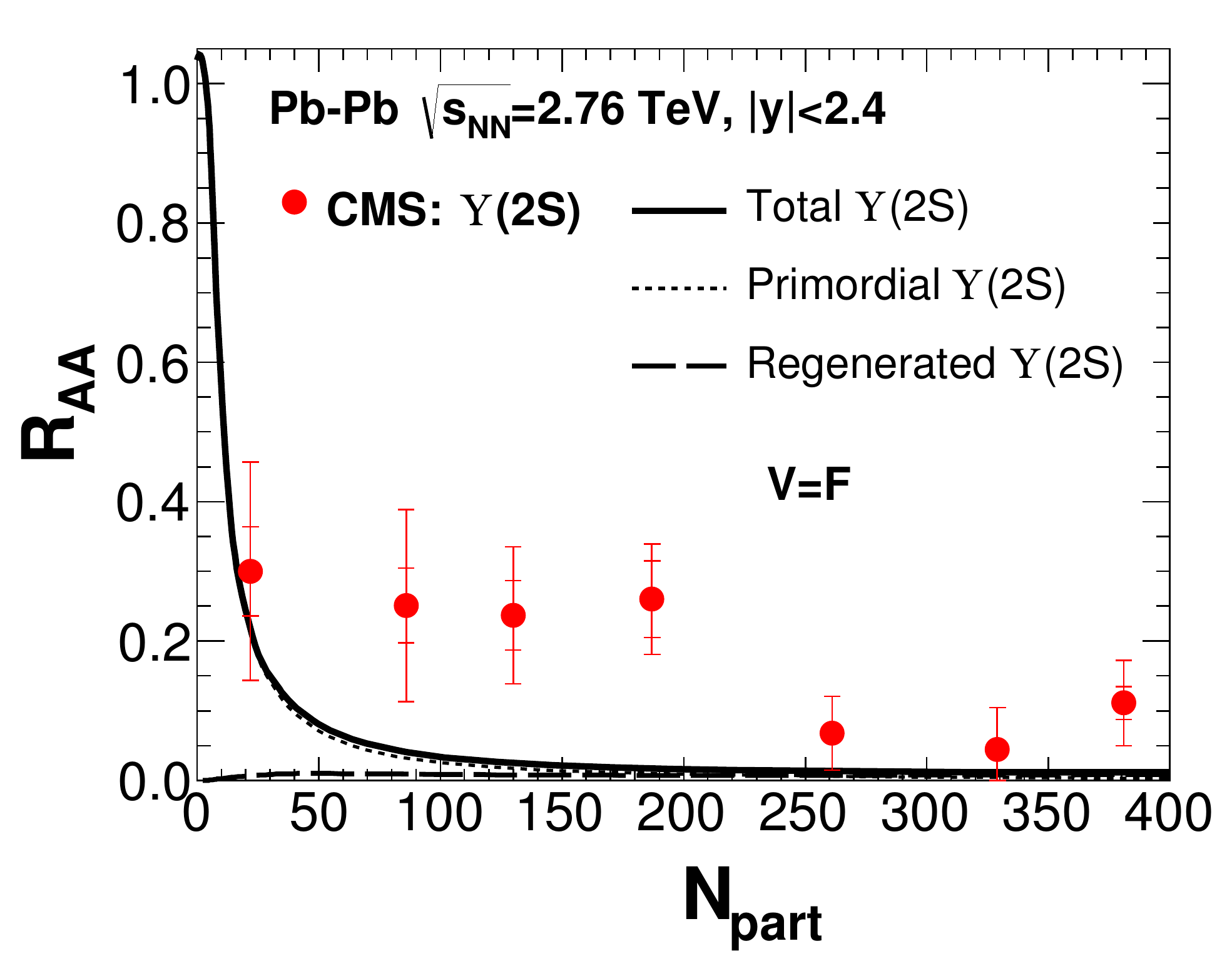}
\caption{The nuclear modification factor for $\Upsilon(2s)$ as a
function of centrality in 2.76 TeV Pb+Pb collisions at mid rapidity. The data are from CMS Collaboration~\cite{cms2}, and the dotted, dashed and solid lines are the transport calculations with only initial production, only regeneration and both in the two limits of $V=U$ (left panel) and $V=F$ (right panel).}
\label{fig2}
\end{figure}

Now we consider the nuclear modification factor $R_{AA}$ for the excited state $\Upsilon(2s)$. The model calculation and the comparison with the data are shown in Fig.\ref{fig2}. Different from the ground state $\Upsilon(1s)$ where there is almost no suppression for the initial production and the small regeneration fraction does not play a key role in the total production, the initially produced $\Upsilon(2s)$s are almost all eaten up in central Pb+Pb collisions, and in this case the small regeneration contribution may become the dominant fraction in the total yield. For $V=U$ the $\Upsilon(2s)$ regeneration happens in a wide temperature region $1 < T/T_c < T_d(\Upsilon(2s))/T_c=1.6$ and the maximum regeneration fraction of $R_{AA}$ reaches 0.1 in most central collisions, see the left panel of Fig.\ref{fig2}. For $V=F$, however, the regeneration region approaches to 0 due to the dissociation temperature $T_d(\Upsilon(2s))/T_c \sim 1$ and there is almost no regeneration, see the right panel of Fig.\ref{fig2}. 
Therefore, only when the heavy quark potential is close to the internal energy, the regeneration becomes the dominant one in the total yield. From the comparison of our transport approach with the CMS data, the data support the calculation with $V=U$.

\begin{figure}[htb]
\centering
\includegraphics[width=0.45\textwidth]{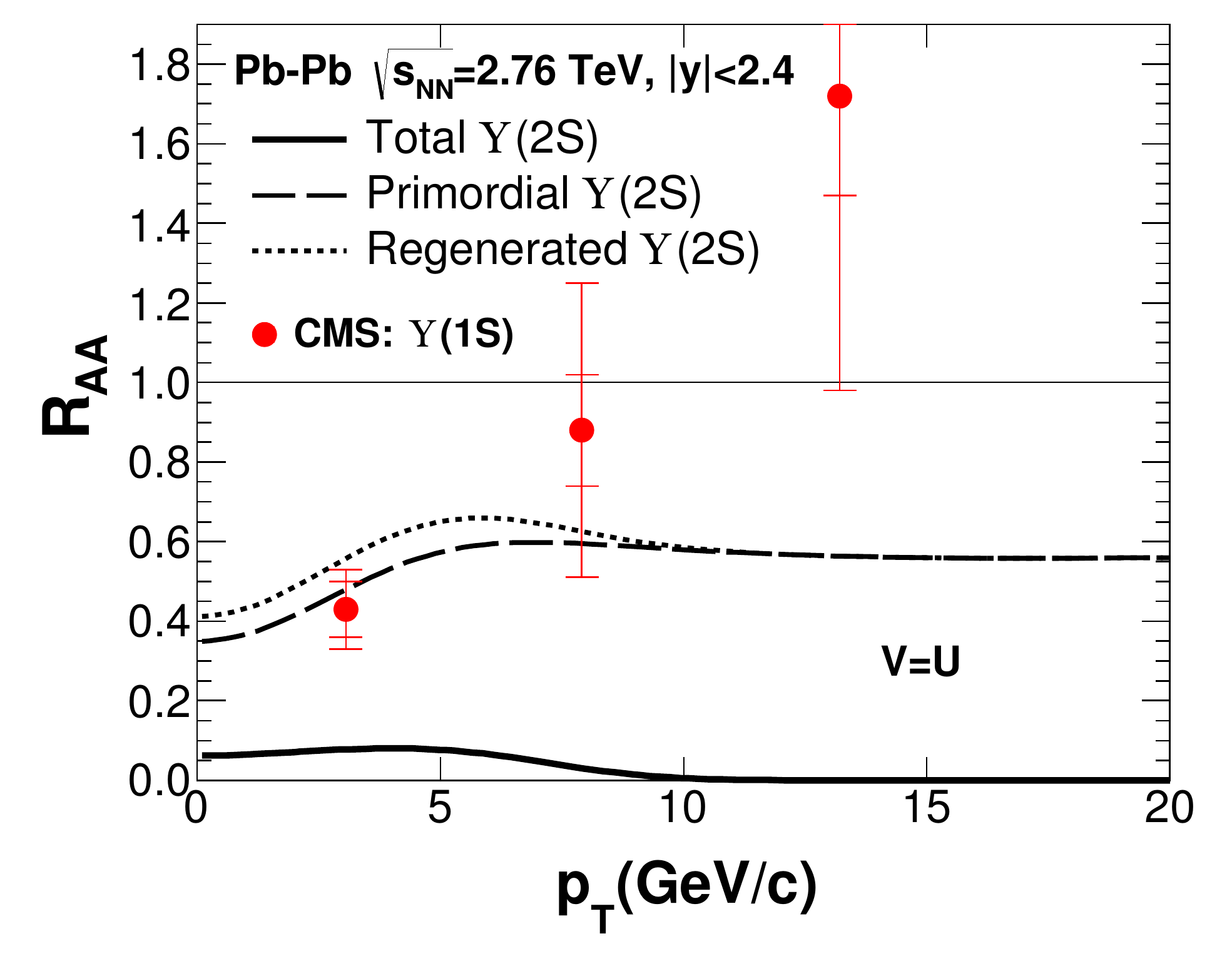}
\caption{The differential nuclear modification factor for $\Upsilon(1s)$ as a
function of transverse momentum in 2.76 TeV Pb+Pb collisions (minimum bias) at mid rapidity. The data are from CMS Collaboration~\cite{cms1}, and the dotted, dashed and solid lines are the transport calculations with only initial production, only regeneration and both in the limit of $V=U$.}
\label{fig3}
\end{figure}

Since the yield is a global quantity, it is not sensitive to the local properties of the hot medium. To look into the dynamical feature of the medium more precisely, we turn to the differential nuclear modification factor $R_{AA}(p_T)$ as a function of transverse momentum $p_T$. Fig.\ref{fig3} shows the calculation for the ground state $\Upsilon(1s)$ with $V=U$ and the comparison with the CMS data in minimum bias Pb+Pb collisions. Since the regeneration happens in the later stage of the fireball evolution, the regenerated quarkonia carry lower momentum and their contribution becomes sizeable only in the low $p_T$ region. The quarkonia with high $p_T$ are all from the initial production. The calculation with $V=U$ agrees well with the data at low $p_T$ but looks underestimated at high $p_T$ where the uncertainty of the data is also very large.

In summary, we investigated the $\Upsilon$ production in heavy ion collisions at LHC in the frame of a dynamical transport approach. The conclusions include: 1) The surviving probability of the excited states is sensitive to the heavy quark potential, and the experimental data support the choice of $V \to U$; and 2) While the regeneration is still small at LHC energy, it controls the nuclear modification factor for the excited states.

\noindent {\bf Acknowledgement:} The work is supported by the NSFC, the MOST, and the DOE grant Nos. 11335005, 2013CB922000, 2014CB845400, and DE-AC03-76SF00098.








\end{document}